\documentclass[twocolumn,prl,preprintnumbers,showpacs,aps,amssymb]{revtex4}

\usepackage{graphicx}
\usepackage{bm}
\usepackage{amsmath}

\def\calL{{\cal L}}
\def\calO{{\cal O}}
\def\calU{{\cal U}}

\def\hbar{{\bar h}}

\def\lambdabar{{\bar\lambda}}

\def\rhohat{{\hat\rho}}
\def\vhat{{\hat v}}

\def\nn{\nonumber}

\begin{document}
\title{Electroweak symmetry breaking from unparticles}
\author{Jong-Phil Lee}
\email{jplee@kias.re.kr}
\affiliation{Korea Institute for Advanced Study, Seoul 130-722, Korea}
\preprint{KIAS-P08021}

\begin{abstract}
A new type of scalar potential inspired by unparticles is proposed for the 
electroweak symmetry breaking.
The interaction between the standard model fields and unparticle sector is
described by the non-integral power of fields that originates from the 
nontrivial scaling dimension of the unparticle operator.
We find that unlike the usual integral-power potential, the electroweak
symmetry is broken at tree level.
The scale invariance of unparticle sector is also broken simultaneously,
resulting in a physical Higgs and a lighter scalar particle.

\end{abstract}
\pacs{11.15.Ex, 12.60.Fr}

\maketitle
{\em Introduction.}---
The secret of electroweak symmetry breaking (EWSB) is a long standing puzzle in
particle physics.
The origin of mass is directly related to EWSB and the resulting gauge hierarchy
problem in the standard model (SM) has been the strongest driving force for
new physics beyond SM.
\par
The existence of a hidden sector can be a good answer for EWSB.
In a minimal extension, the hidden sector scalar couples to the SM scalar field
in a scale-invariant way \cite{Meissner,Espinosa,CNW,FKV}.
It is quite well known that the scale invariance is broken at quantum loop 
level, and the scalar field achieves the vacuum expectation value (VEV) through
the Coleman-Weinberg (CW) mechanism \cite{CW}.
Since in the original CW mechanism the mass scale is generated radiatively 
with the conformal symmetry breaking, the Higgs mass is much smaller 
($\lesssim10$ GeV) than the LEP bound ($\gtrsim115$ GeV).
The additional scalar field from the hidden sector evades this difficulty,
and can provide a good candidate for dark matter.
\par
Recently the hidden sector has received much attention with the possibility 
of the existence of unparticles \cite{Georgi}.
The unparticle is a scale invariant stuff in a hidden sector.
Its interactions with the SM particles are well described by an effective
theory formalism.
\par
The most striking feature of the unparticle is its unusual phase space with
non-integral scaling dimension $d_\calU$.
For an unparticle operator of scaling dimension $d_\calU$, 
the unparticle appears as a non-integral number $d_\calU$ of invisible 
massless particles.
After the Georgi's suggestion, there have been a lot of phenomenological
studies on unparticles \cite{Cheung,Fox,DEQ,Stephanov,jplee}.
\par
In this Letter, we investigate the possibility of EWSB from unparticles.
The framework is very similar to that of EWSB with hidden sector scalar fields,
but the hidden scalar sector is replaced by the scalar unparticle sector.
Among other couplings between SM fields and unparticles, Higgs-unparticle
interaction is very interesting because its coupling is relevant \cite{Fox};
\begin{equation}
\calL_{\Phi\Phi\calU}\sim 
\lambda_{\Phi\Phi\calU}(\Phi^\dagger\Phi)\calO_\calU~,
~~~[\lambda_{\Phi\Phi\calU}]=2-d_\calU>0~,
\end{equation}
where $\Phi$ is a fundamental Higgs, $\calO_\calU$ is a scalar unparticle 
operator with scaling dimension $1<d_\calU<2$, 
$\lambda_{\Phi\Phi\calU}$ is the coupling constant, 
and $[~\cdot~]$ calculates the mass dimension.
\par
The main motivation of this work is the observation that the scalar unparticle
operator $\calO_\calU$ is {\em equivalent} to $d_\calU$ number of massless
particles.
We propose a new type of scalar potential
\begin{equation}
V_{int}\sim\lambda(\Phi^\dagger\Phi)(\phi^*\phi)^{d_\calU/2}~,
\end{equation}
where $\phi$ is a massless scalar field with $[\phi]=1$.
Note that the usual scalar potential for EWSB from hidden scalar sector 
contains the marginal interaction term of $\lambda|\Phi|^2|\phi|^2\subset V_0$.
With the quartic terms of $\Phi$ and $\phi$, it can be shown that there is some
ray of fields in $V_0$ along which $V_0$ has nontrivial minimum equal to the
trivial minimum value $V_0(0)=0$ \cite{GW}.
When the radiative corrections are turned on, there appears a small curvature
along the radial direction and the VEV is picked out.
Since there is no scale at tree level for $V_0$, this is a typical example of
the dimensional transmutation.
\par
On the contrary, if one considers the scalar potential containing the form of
$V_{int}$, it inevitably introduces a mass scale through the dimensionful
coupling.
One may expect that there is a nontrivial minimum along the radial direction
{\em at tree level} for $V\supset V_{int}$.
It will be shown that this is indeed the case.
In other words, interactions between the SM fields and unparticle sector
themselves break the electroweak symmetry.
\par
When EWSB occurs one expands the scalar fields around the vacuum.
The resulting fluctuations mix up with each other to form two physical scalar
states.
In this simple setup, it is quite natural to identify a heavy state as Higgs.
The other light state has a mass proportional to $(2-d_\calU)$ which vanishes
as $d_\calU\to 2$.
This is the remnant of the fact that $V_0$ has a massless scalar at tree level
as a pseudo Goldstone boson from the conformal symmetry breaking.
The unparticle sector thus no longer remains scale-invariant after the EWSB.
So the interaction $V_{int}$ induces {\em both} EWSB in the SM sector and the
scale-invariance breaking in the unparticle sector.
We find that all of these things can happen for acceptable values of the 
parameters of this setup.
\par
The Letter is composed as follows. 
In the next section, the new potential is proposed and its properties are
investigated.
The way of how the EWSB occurs is also given.
After that, the resulting mass spectrum is analyzed.
The concluding remarks appear at the end.
{\em Scalar Potential.}---
We start with the scalar potential of the form
\begin{equation}
V(\Phi,\phi)=\lambda_0(\Phi^\dagger\Phi)^2+\lambda_1(\phi^*\phi)^2
  +2\lambda_2\mu^{2-d_\calU}(\Phi^\dagger\Phi)
 (\phi^*\phi)^{{d_\calU}/{2}}~,
\label{V}
\end{equation}
where $\lambda_0$ is assumed to be positive. Here the mass
dimension of $\phi$ is $1$ and a dimension-1 parameter $\mu$ is
inserted to make $\lambda_2$ dimensionless. As in \cite{GW}, we
try to find the minimum of $V$ along some ray $\Phi_i=\rho N_i$,
where $\vec{N}$ is a unit vector in the field space
$\Phi_i=(\Phi,\phi)$. In unitary gauge, the fields are
parameterized as
\begin{equation}
\Phi=\frac{\rho}{\sqrt{2}}
\left(\begin{array}{c}
0\\N_0\end{array}\right)~,~~~
\phi=\frac{\rho}{\sqrt{2}}N_1~,
\end{equation}
where $N_0^2+N_1^2=1$.
The scalar potential becomes
\begin{equation}
V(\rho,{\vec N})=\frac{\rho^4}{4}\left[
\lambda_0N_0^4+\lambda_1N_1^4
+\left(\frac{\rhohat^2}{2}\right)^{-\epsilon}
 2\lambda_2N_0^2N_1^{d_\calU}\right]~,
\end{equation}
where $d_\calU\equiv 2-2\epsilon$, and $\rhohat\equiv\rho/\mu$.
For $1<d_\calU<2$, one has $0<\epsilon<1/2$.
\par
The stationary condition for $V$ along the $\vec N$ direction for some
specific unit vector ${\vec N}={\vec n}$,
$(\partial V/\partial N_i)_{\vec n}=0$, gives
\begin{eqnarray}
\left(\frac{\rhohat^2}{2}\right)^{-\epsilon}\lambda_2 n_1^{d_\calU}&=&
-\lambda_0n_0^2~,\\
2\lambda_1n_1^4&=&d_\calU\lambda_0 n_0^4~.
\end{eqnarray}
Combining the normalization of $\vec n$ ($n_1^2+n_2^2=1$), one gets
\begin{eqnarray}
\label{rho}
n_0^2&=&\frac{\sqrt{2\lambda_1}}
 {\sqrt{d_\calU\lambda_0}+\sqrt{2\lambda_1}}~,\nn\\
n_1^2&=&\frac{\sqrt{d_\calU\lambda_0}}
 {\sqrt{d_\calU\lambda_0}+\sqrt{2\lambda_1}}~.
\end{eqnarray}
In order for $V$ to have a minimum at $\vec N=\vec n$,
its second derivative must be non-negative.
For any vector $u_i$, one can easily find that
\begin{equation}
\frac{\partial^2 V}{\partial N_i\partial N_j}\Bigg|_{\vec n} u_iu_j\ge 0~.
\end{equation}
In case of $d_\calU=2$, $V(\vec n)=0=V(\rho=0)$, irrespective of $\rho$.
To get a nontrivial minimum along $\rho$, the CW mechanism is implemented.
But if $1<d_\calU<2$,
\begin{equation}
V(\rho, {\vec n})=\frac{\rho^4}{4}\lambda_0n_0^4(-\epsilon)
<0=V(\rho=0)~.
\end{equation}
One interesting point is that the value of $\rho$ is fixed by the
$\vec N$-stationary condition, as given in Eq.\ ({\ref{rho}):
\begin{equation}
\rho=\rho_0\equiv\Bigg(
 -\frac{2^\epsilon\lambda_2 n_1^{d_\calU}}{\lambda_0n_0^2}
 \Bigg)^{\frac{1}{2\epsilon}}\mu~.
\label{rho0}
\end{equation}
One can also easily
find that at $\rho=\rho_0$ along ${\vec n}$,
\begin{equation}
\frac{\partial V}{\partial\rho}\Bigg|_{\rho_0,\vec n}=0~,~~~
\frac{\partial^2 V}{\partial\rho^2}\Bigg|_{\rho_0, \vec n}
=2\epsilon\rho_0^2(\lambda_0 n_0^4+\lambda_1 n_1^4)>0~.
\end{equation}
In short, we have found a minimum of the scalar potential {\em at tree level}
by combining the scalar unparticle sector with the SM scalar field.
\par
It should be noted that when $d_\calU\to2$, $\rho_0$ goes to $0$ or infinity
depending on the values of $\lambda_{0,2}$ and $n_{0,1}$.
Since the vacuum expectation value of $\rho$ is directly proportional to the
mass scale of the theory (e.g., gauge boson masses, Higgs masses, etc.),
it is not desirable if $\rho_0$ gets too small or too large for $d_\calU\to2$.
We require that $\rho_0$ is stable for $d_\calU\to2$ ($\epsilon\to0$).
A little algebra shows that this requirement is satisfied if
\begin{equation}
\lambda_2=-\sqrt{\lambda_0\lambda_1}\equiv\lambdabar~.
\end{equation}
In fact, $\lambdabar$ is the value of $\lambda_2$ for $d_\calU=2$
\cite{CNW}. For $\lambda_2=\lambdabar$ one has
\begin{eqnarray}
\rhohat_0^2&=&
2\left(\frac{d_\calU}{2}\right)^{\frac{1}{2-d_\calU}}
\frac{\sqrt{d_\calU\lambda_0}+\sqrt{2\lambda_1}}{\sqrt{d_\calU\lambda_0}}\nn\\
&\to&
 \frac{2}{\sqrt{e}}
\frac{\sqrt{\lambda_0}+\sqrt{\lambda_1}}{\sqrt{\lambda_0}}~,
~~~{\rm as}~~~d_\calU\to2~.
\label{rhohat0}
\end{eqnarray}
But when $d_\calU=2$, $\rho_0$ is no longer a global minimum and
$\rho$ cannot develop the vacuum expectation value at tree level.
\par
{\em Mass Spectrum.}---
When $\lambda_{1,2}$ are turned on, the potential $V$ develops the VEV
at $\rho=\rho_0$. 
Around $v$ the fields $\Phi$
and $\phi$ are expanded with fluctuations $h$ and $s$ as
\begin{equation}
\Phi=\frac{1}{\sqrt{2}}\left(\begin{array}{c}
0\\n_0\rho_0+h\end{array}\right)~,~~~
\phi=\frac{1}{\sqrt{2}}(n_1\rho_0+s)~.
\end{equation}
The scalar potential now becomes
\begin{eqnarray}
V(h,s)&=&
\frac{\lambda_0}{4}(n_0\rho_0+h)^4+\frac{\lambda_1}{4}(n_1\rho_0+s)^4\nn\\
&&+2^{-d_\calU/2}\lambda_2\mu^{2\epsilon}
(n_0\rho_0+h)^2(n_1\rho_0+s)^{d_\calU}~.
\end{eqnarray}
The mass squared matrix for $h$ and $s$ is
\begin{eqnarray}
(M^2)_{i,j}&=&\frac{\partial^2V}{\partial\psi_i\partial\psi_j}\Bigg|_0\nn\\
&=&
\frac{\rho_0^2n_0^2}{\sqrt{2\lambda_1}}
\left(\begin{array}{cc} 2\lambda_0\sqrt{2\lambda_1} &
-\left(2d_\calU\lambda_0\lambda_1\right)^\frac{3}{4}\\
-\left(2d_\calU\lambda_0\lambda_1\right)^\frac{3}{4} &
(4-d_\calU)\lambda_1\sqrt{d_\calU\lambda_0}\end{array}\right)~,\nn\\
\end{eqnarray}
where $\psi_i=(h,s)$.
Two eigenvalues of $M^2$ correspond to the heavy and light scalar
mass squared as follows:
\begin{equation}
m_{h,\ell}^2= \frac{\rho_0^2\sqrt{2\lambda_0\lambda_1}}
{\sqrt{d_\calU\lambda_0}+\sqrt{2\lambda_1}}\left\{
\sqrt{\lambda_0}+\left(2-\frac{d_\calU}{2}\right)\sqrt{\frac{d_\calU}{2}\lambda_1}
\pm\sqrt{D}\right\}~,
\end{equation}
where
\begin{equation}
D=\lambda_0+\left(2-\frac{d_\calU}{2}\right)^2\frac{d_\calU}{2}\lambda_1
+\left(\frac{3d_\calU}{2}-2\right)\sqrt{2d_\calU\lambda_0\lambda_1}~.
\end{equation}
For a small $\epsilon=1-d_ \calU/2\ll 1$,
\begin{eqnarray}
\frac{m_h^2}{\rho_0^2}&=&2\sqrt{\lambda_0\lambda_1}\left[
1+\frac{\epsilon}{2}\left(\frac{\sqrt{\lambda_0}-\sqrt{\lambda_1}}
{\sqrt{\lambda_0}+\sqrt{\lambda_1}}\right)^2\right]~,\\
\frac{m_\ell^2}{\rho_0^2}&=&
2\sqrt{\lambda_0\lambda_1}\left[\frac{2\epsilon\sqrt{\lambda_0\lambda_1}}
{(\sqrt{\lambda_0}+\sqrt{\lambda_1})^2}\right]~.
\end{eqnarray}
Note that the value of $2\sqrt{\lambda_0\lambda_1}\rho_0^2$ is the
heavy scalar mass squared for $d_\calU=2$, and is identified with
the Higgs mass squared \cite{CNW}. We also identify $m_h$ as Higgs
boson, and $m_\ell$ as a new light scalar.
\par
When $d_\calU=2$, the light scalar is massless at tree level. 
The reason is that it corresponds to the pseudo Goldstone boson from
the spontaneous symmetry breaking of the conformal symmetry 
\cite{GW,Goldberger}. 
The light scalar boson is called the "scalon." The scalon gets massive
by the CW mechanism.
\par
But for $\epsilon=1-d_\calU/2\ll 1$, we have found that
$m_\ell^2/m_h^2\sim\epsilon$ at tree level. Thus the new light
scalar and Higgs boson masses are good probes to the hidden
unparticle sector.
\par
The vacuum expectation value $\rho_0$ is related to the gauge
boson ($W$) masses:
\begin{equation}
m_W^2=\frac{1}{4}g_W^2(n_0\rho_0)^2=\frac{\sqrt{2}g_W^2}{8G_F}~,
\end{equation}
where $g_W$ is the weak coupling and $G_F$ is the Fermi constant.
Thus we can fix
\begin{equation}
(n_0\rho_0)^2=\frac{1}{\sqrt{2}G_F}=(246~{\rm GeV})^2\equiv
v_0^2~.
\end{equation}
Combining Eq.\ (\ref{rhohat0}) yields
\begin{equation}
\vhat_0^2=2\left(\frac{d_\calU}{2}\right)^{\frac{d_\calU}{2(2-d_\calU)}}
\sqrt{\frac{\lambda_1}{\lambda_0}}~.
\label{vhat0}
\end{equation}
where $\vhat_0=v_0/\mu$. 
The right-hand-side of Eq.\ (\ref{vhat0}) is a slow varying function of 
$d_\calU$. 
If one chooses $\mu=v_0$, the ratios of couplings are
\begin{eqnarray}
\frac{\lambda_1}{\lambda_0}
&=&\frac{1}{4}\left(\frac{2}{d_\calU}\right)^{\frac{d_\calU}{2-d_\calU}}
\longrightarrow\frac{e}{4}\simeq 0.68~~{\rm as}~d_\calU\to 2~, \nn\\
\frac{\lambda_2}{\lambda_0}&=&-\sqrt{\frac{\lambda_1}{\lambda_0}}
\longrightarrow-0.82~.
\label{fixratio}
\end{eqnarray}
When $d_\calU=1$, $\lambda_1/\lambda_0=0.5$ and 
$\lambda_2/\lambda_0\simeq-0.71$.
Since the ratios are of order 1 for all range over $d_\calU$, 
the scale of $\mu$ around the weak scale is a reasonable choice. 
In other words, interactions
between the SM sector and unparticle sector at the electroweak
scale are quite plausible.
\begin{figure}
\includegraphics{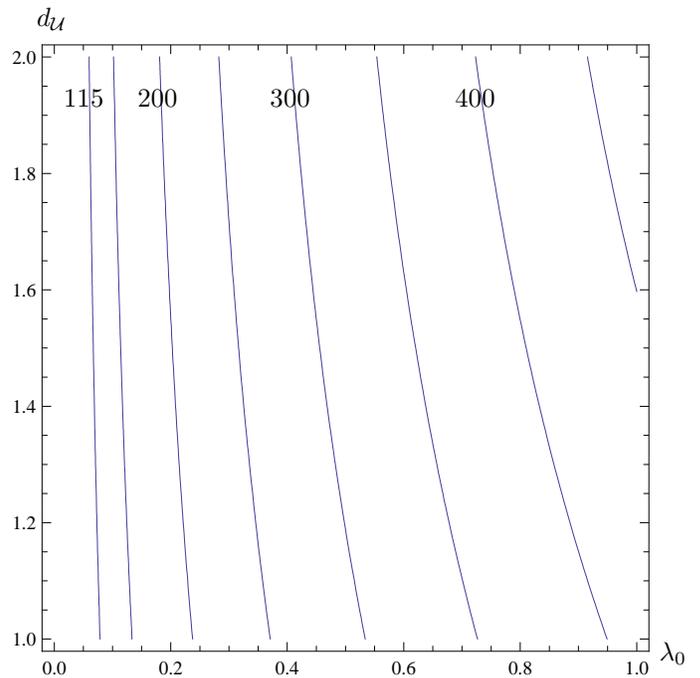}
\caption{\label{contour}Plots of $(\lambda_0,d_\calU)$ with the fixed ratios of
Eq.\ (\ref{fixratio})
for $m_h=$ 115, 150, 200, 250, 300, 350, 400, 450 (GeV), from left to right.}
\end{figure}
\begin{figure}
\includegraphics{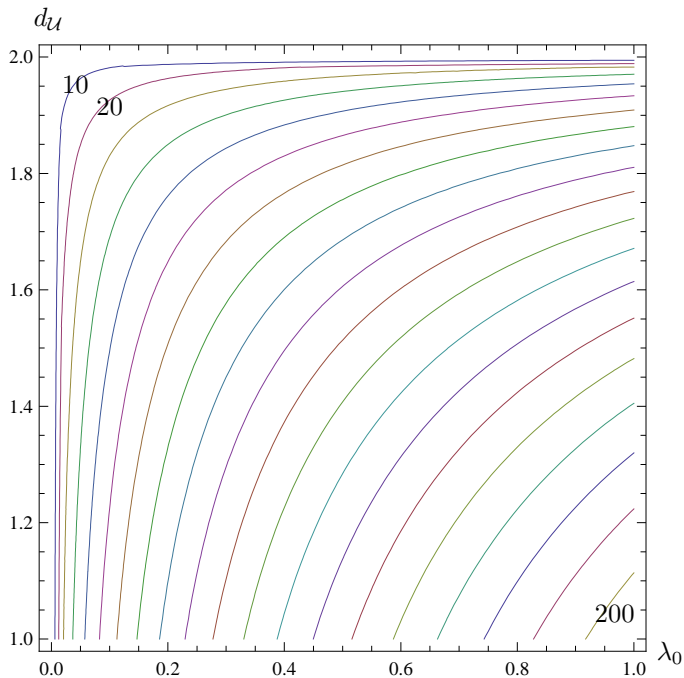}
\caption{\label{mLctr}Plots of $(\lambda_0,d_\calU)$ with the fixed ratios of
Eq.\ (\ref{fixratio})
for $m_\ell=$ 10, 20, 
$\cdots$, 200 (GeV).}
\end{figure}
Figure \ref{contour} (\ref{mLctr}) shows the possible values of 
$(\lambda_0,d_\calU)$ for
various Higgs (light scalar) masses.
One can see that heavier scalars constrain $d_\calU$ more strongly.
Thus the discovery of, say, Higgs alone of mass $\lesssim 400$ GeV will not 
give much information on $d_\calU$.
\begin{figure}
\includegraphics{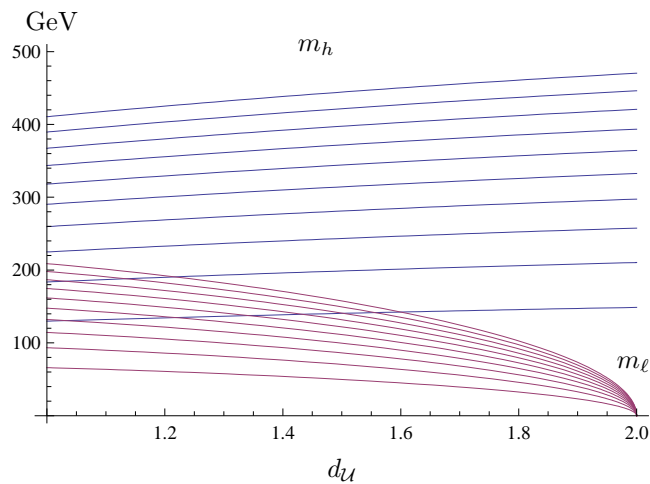}
\caption{\label{mass}Scalar masses $m_h$ and $m_\ell$ as a function of $d_\calU$
with the fixed ratios of Eq.\ (\ref{fixratio})
for $\lambda_0=$ 0.1, 0.2, $\cdots$, 1.0, from bottom to top.}
\end{figure}
As also given in Fig.\ \ref{mass}, $m_h$ is rather inert with respect to 
$d_\calU$ while $m_\ell$ is not.
With the condition of Eq.\ (\ref{fixratio}), both $m_{h,\ell}$ are proportional
to $\sim\sqrt{\lambda_0}$.
One can find that
\begin{eqnarray}
130(149)~{\rm GeV}&\lesssim& m_h\lesssim 411(470)~{\rm GeV}~,\nn\\
66~{\rm GeV}&\lesssim& m_\ell\lesssim 209~{\rm GeV}~,
\label{massrange}
\end{eqnarray}
for $d_\calU=1(2)$.
As $d_\calU$ increases $m_h$ increases slightly while $m_\ell$ decreases and
finally vanishes at $d_\calU=2$, 
and the gap between $m_h$ and $m_\ell$ gets larger as $d_\calU$ increases.
If the scalar masses turned out to be quite different from 
Eq.\ (\ref{massrange}), then the value of $\mu$ should be rearranged to fit
the data.
But in this case one would have to explain why that value of $\mu$ is so
different from $v_0$, the electroweak scale.

{\em Conclusions.}---
In this Letter we suggest a new scalar potential with a fractional power of
fields from hidden sector inspired by the scalar unparticle operator.
Unlike the usual potential of marginal coupling, the new one develops VEV at
tree level.
In this picture, the EWSB occurs when the unparticle sector begins to interact
with the SM sector.
If the hidden sector were not scale invariant and the coupling were marginal,
the EWSB happens radiatively through the CW mechanism.
When the scaling dimension $d_\calU$ departs from the value of $2$ 
a new scale (of the order of $\sim 1/\sqrt{G_F}$) is introduced in the scalar 
potential through the relevant coupling,
and the electroweak symmetry is broken at tree level.
In other words, the EWSB occurs when the hidden sector enters the regime of 
scale invariance, i.e., unparticles.
In view of the unparticle sector, the new potential also breaks the scale
invariance of the hidden sector.
\par
Once the electroweak symmetry is broken, the scalar fields from SM and hidden
sector mix together to form two massive physical states.
The heavy one is identified as Higgs, while the light one is a new particle
of mass around $\lesssim 230$ GeV.
The possibility of the new light scalar to be a dark matter will be a good
challenge for future studies.
\par
If the hidden sector self coupling $\lambda_1$ vanishes, then the minimum of
the potential appears along the ray of $\Phi=0$.
In this case the VEV cannot produce the $W$ boson mass $m_W$ since $m_W$ 
occurs when the fluctuation is transverse to the $\Phi=0$ direction.
Thus our potential $V(\Phi,\phi)$ in Eq.\ (\ref{V}) is minimal.


\end{document}